\documentclass[intlimits,twoside,a4paper]{article}
\usepackage[cp1251]{inputenc}

\usepackage[eqsecnum]{cmpj3}

\usepackage{bm}

\issue{2018}{21}{3}{33706}
\doinumber{10.5488/CMP.21.33706}

\title[$t$-$J$ model on the honeycomb lattice]{Electronic spectrum and superconductivity \\ in the $t$-$J$ model on the honeycomb lattice%
\thanks{The paper is devoted to the 80-th
birthday of Professor I.V.~Stasyuk.}}
\author[N.M. Plakida]{N.M. Plakida}
\address{Joint Institute for Nuclear Research, 141980 Dubna, Russia }

\begin{document}
	
\date{Received June 1, 2018, in final form July 27, 2018}

\maketitle

\begin{abstract}
A microscopic theory of electronic  spectrum and superconductivity
 within the $t$-$J$ model on the honeycomb lattice is formulated. The Dyson
equation for the normal and anomalous Green functions for the two-band model in terms of
the Hubbard operators is derived by applying the Mori-type projection technique.  The
self-energy is evaluated in the self-consistent Born approximation for electron
scattering on spin and charge  fluctuations induced by the kinematical  interaction for
the Hubbard operators.  Superconducting pairing mediated by the antiferromagnetic
exchange and spin fluctuations is discussed.
\keywords strong electron correlations, $t$-$J$ model, honeycomb lattice,
superconductivity
\pacs 74.20.Mn, 71.27.+a, 71.10.Fd, 74.72.-h
\end{abstract}

\section{Introduction}
\label{sec:1}

One of the crucial issues in the current theory of condensed matter is the study of
superconductivity  in strongly correlated electronic systems, as in the cuprate
high-temperature superconductors (see, e.g., ~\cite{Plakida10}). The basic model commonly
used in these studies is the Hubbard model which is specified by  two parameters:  the
single-electron hopping matrix element $t$  and the single-site Coulomb energy $U$. A
rigorous analytical treatment of the model is based on the Hubbard operator (HO)
technique~\cite{Hubbard65} since in this representation the local constraint of no double
occupancy of any lattice site is rigorously implemented by the Hubbard operator algebra.
A superconduction pairing due to the kinematical interaction in the Hubbard model in the
limit of strong electron correlations ($U \to \infty $) was first considered by Zaitsev
and Ivanov~\cite{Zaitsev87} who used  a diagram technique for the HOs. Subsequently,
superconducting pairing in the Hubbard model was studied by Plakida and
Stasyuk~\cite{Plakida88} by applying the equation of motion method for the thermodynamic
Green functions (GFs)~\cite{Zubarev60,Zubarev74}. In the later studies ~\cite{Plakida89,Yushankhai91}
of the $t$-$J$ model within
the Mori-type projection technique~\cite{Mori65,Plakida11}, a formula for the
superconduction temperature similar to~\cite{Zaitsev87} was obtained in the mean-field
approximation (MFA).  A self-consistent solution of  Dyson equations beyond the MFA for
normal and anomalous GFs and respective self-energies within  the $t$-$J$ model was
performed in~\cite{Plakida99} that confirmed the  results of the previous studies in
MFA~\cite{Plakida89,Yushankhai91}. Detailed studies of the extended Hubbard model on the
square lattice  were performed in~\cite{Plakida13,Plakida14}. Taking into account an
intersite Coulomb repulsion and electron-phonon interaction,  a microscopic theory of
superconductivity in cuprates was formulated. It was proved that the spin-fluctuation
pairing mechanism induced by the strong kinematical interaction is the leading
interaction which results in high-temperature superconductivity.
\par
 In recent years, the two-dimensional carbon honeycomb lattice, the graphene,   has been
extensively  studied due its peculiar electronic properties (for a review
see~\cite{Neto09,Kotov12}). Studies of the graphene beyond the simple  model of
noninteracting electrons  by taking into account the Coulomb interaction  reveal a rich
phase diagram with phase transitions to the antiferromagnetic (AF) state, spin-density
wave (SDW), charge-density wave (CDW), and  nonconventional superconductivity (SC).
Superconducting phase transitions in the Hubbard model on the  honeycomb lattice have
been considered in several publications. The renormalization group  approach was used
in~\cite{Honerkamp08} to study phase transitions in the extended Hubbard model with  a
moderate on-site interaction $U$. Close to half-filling, the SDW or CDW orders occur,
while  for a large doping,  $f$-wave triplet-pairing and $d + \ri d$-wave singlet-pairing
appear. Using the dynamic cluster approximation for the Hubbard model with  $U/t = 2 -
6$, a transition from the $d + \ri d$-wave singlet pairing at weak coupling  to the $p$-wave
triplet pairing at larger coupling was observed in~\cite{Xu16}.

In the limit of strong correlations, $U \gg t$, two conduction bands of electrons in the
honeycomb lattice  split   into the singly- and doubly-occupied Hubbard subbands. In this
limit, the Hubbard model can be reduced to the two-band $t$-$J$ model for the projected
electron operators.  A detailed study of the $t$-$J$ model on the honeycomb lattice was
presented in~\cite{Gu13}. The ground-state energy and the staggered magnetization in the
AF phase as function of doping $\delta$ have been calculated using the Grassmann tensor
product state approach, exact diagonalization and density-matrix renormalization methods.
The occurrence of the time-reversal symmetry breaking $d + \ri d$-wave SC at large doping
was found. Moreover, a coexisting of the  SC and AF order was observed for low doping, $
0 <\delta< 0.1$, where the triplet pairing is induced.

In the above cited  papers, mostly the phase diagram of the   models with the Coulomb
interaction on the honeycomb lattice was studied at zero temperature by numerical
methods. To investigate temperature and doping dependence of the phase diagram and, in
particular, superconduction phase transition  analytical methods should be used. The
honeycomb Heisenberg model at half-filling over the whole temperature region both in the
AF  and paramagnetic phases was studied in~\cite{Vladimirov17}.  The electronic and
spin-fluctuation spectra in this model were studied in~\cite{Vladimirov18}  using the
generalized mean-field approximation (GMFA). In the present paper, we formulate a
microscopic theory of superconductivity within the two-band  $t$-$J$ model on the
honeycomb lattice. We derive Dyson equation for the normal and anomalous matrix GFs with
the self-energy evaluated in the self-consistent Born approximation (SCBA) as for the
one-band $t$-$J$ model  in~\cite{Plakida99} and the Hubbard model
in~\cite{Plakida13,Plakida14} on the square lattice. Electronic spectrum and
superconductivity is  considered  in the MFA.

In section~\ref{sec:2}, we formulate the $t$-$J$  model in terms of  Hubbard  operators.
The Dyson equation is derived in  section~\ref{sec:3}.  The results of calculation of the
electronic spectrum and the superconduction gap function are presented in
section~\ref{sec:4}.  The conclusion can be found in section~\ref{sec:5}.

\section{The $t$-$J$ model }
\label{sec:2}

We consider the  $t$-$J$  on the honeycomb lattice. The honeycomb lattice is bipartite
with two triangular sublattices A and B. Each of the $N$ sites on the A sublattice
is connected to three nearest-neighbor (nn) sites belonging to the B sublattice by
vectors ${{\boldsymbol{\delta}}_\alpha}$, and $N$ sites on B are connected to A by vectors $-{{\boldsymbol{\delta}}_\alpha}$:
\begin{equation}
{{\boldsymbol{\delta}}_1} = \frac{a_0}{2}\big(\sqrt{3}, -1\big)\,,\; {{\boldsymbol{\delta}}_2} = -
\frac{a_0}{2}\big(\sqrt{3}, 1\big)\,,\; {{\boldsymbol{\delta}}_3} = a_0(0, 1). \label{nn}
\end{equation}
The basis vectors  are ${\bf a}_1 = {{\boldsymbol{\delta}}_3} - {{\boldsymbol{\delta}}_2} =
({a_0}/{2})(\sqrt{3}, 3)$ and ${\bf a}_2 = {{\boldsymbol{\delta}}_3} - {{\boldsymbol{\delta}}_1} =
({a_0}/{2})(-\sqrt{3}, 3)$, the lattice constant is $a = |{\bf a}_1| = |{\bf a}_2| =
\sqrt{3}a_0$, where $a_0$ is  the nn distance; hereafter we put $a_0 = 1$.  The
reciprocal lattice   vectors are ${\bf k}_1 = ({2\piup}/{3})(\sqrt{3}, 1)$ and $ {\bf k}_2
=({2\piup}/{3}) (-\sqrt{3}, 1)$.

 In conventional notation, the  $t$-$J$ model reads:
\begin{equation}
H =  - t\, \sum_{\langle i,j \rangle \sigma}  \tilde a_{i,\sigma}^{+} \tilde a_{j,\sigma}
- \mu \sum_{i , \sigma} \,n_{i,\sigma} + \frac{J}{2}\sum_{\langle i, j \rangle}\;\left(
{\bf S}_{i} \; {\bf S}_{j}
 - \frac{1}{4} n_i\, n_j \right),
\label{b1}
\end{equation}
where for the two-sublattice representation A, B  of the site indices on the honeycomb
lattice, a shorter notation   $ i \alpha \rightarrow i $ is used. The projected electron
operators in the singly occupied Hubbard subband in the model (\ref{b1}) are defined as
$\tilde a_{i,\sigma}^{+} =a_{i,\sigma}^{+}(1-n_{i ,\bar{\sigma}})\;$ and $ \tilde
a_{i\sigma} =a_{i\sigma}(1-n_{i ,\bar{\sigma}})$ for creation and annihilation of an
electron  with spin $\sigma/2$ ($\sigma=\pm 1 , \; \bar{\sigma} = - \sigma$) , $n_{i
,\sigma}  = \tilde a_{i,\sigma}^{+}\, \tilde a_{i,\sigma}$, and $n_i = \sum_{ \sigma}
n_{i,\sigma}$. Here, $t$ is the nn electron hopping energy  and $J = 4t^2/U$ is the  nn
AF exchange interaction for electronic spins
 ${\bf S}_{i}$.

To take into account, on a rigorous basis, the no-double-occupancy constraint for the
projected electron operators $\tilde a_{i,\sigma}^{+} $, we employ the Hubbard operator
(HO) technique~\cite{Hubbard65}. The HOs  $\, X_{i}^{nm}=|i,n\rangle\langle i,m| \,$
determine transitions between three possible states on a lattice site $i$:
$|i,n\rangle=|i,0\rangle$  and $|i,\sigma\rangle$ for an empty site and for a singly
occupied site by an electron with spin $\sigma/2$, respectively. The electron number
operator and the spin operators in terms of HOs are defined as
\begin{eqnarray}
  n_i &=& \sum_{\sigma} X_{i}^{\sigma \sigma} =   X_{i}^{++}  +  X_{i}^{--},
\label{b3a}\\
S_{i}^{\sigma} &=& X_{i}^{\sigma\bar\sigma} ,\quad
 S_{i}^{z} =  (\sigma/2) \,[ X_{i}^{\sigma \sigma}  -
  X_{i}^{\bar\sigma \bar\sigma}] .
\label{b3b}
\end{eqnarray}
The completeness relation for the HOs, $\, X_{i}^{00} +  \sum_{\sigma}
X_{i}^{\sigma\sigma}  = 1 $, rigorously preserves the constraint of no-double-occupancy
for a quantum state $|i,n\rangle $ on any lattice site $i$. From the multiplication rule
$\, X_{i}^{nm} X_{i}^{kl} = \delta_{mk} X_{i}^{nl} \,$ for HOs, there follow the commutation
relations:
\begin{equation}
\left[X_{i}^{nm}, X_{j}^{kl}\right]_{\pm}= \delta_{ij}\left(\delta_{mk}X_{i}^{nl}\pm
\delta_{nl}X_{i}^{km}\right).
 \label{b4}
\end{equation}
The upper sign  refers to  Fermi-type operators such as  $X_{i}^{0\sigma}$, while the
lower sign refers to  Bose-type operators such as  $n_i$  (\ref{b3a}) or the spin
operators (\ref{b3b}).
\par
Using the Hubbard operator representation  for $\tilde a_{i\sigma}^{+} = X_{i}^{\sigma 0}$\,\,,  $\tilde a_{j\sigma}= X_{j}^{0\sigma}$ and equations~(\ref{b3a}) and~(\ref{b3b}),
we  write the Hamiltonian of the $t$-$J$ model (\ref{b1}) in the form:
\begin{eqnarray}
H = - t\sum_{\langle i, j \rangle \sigma}\,  X_{i}^{\sigma 0}\, X_{j}^{0\sigma}
 - \mu \sum_{i \sigma} X_{i}^{\sigma \sigma}
  +\frac{J}{4} \sum_{\langle i, j \rangle \sigma}\,
\left(X_i^{\sigma\bar{\sigma}}X_j^{\bar{\sigma}\sigma}  -
   X_i^{\sigma\sigma}X_j^{\bar{\sigma}\bar{\sigma}}\right).
\label{b5}
\end{eqnarray}
The chemical potential $\mu$ depends on the average electron occupation number $n = n_\text{A}=
n_\text{B}$\,,
\begin{equation}
  n_{\alpha} = \frac{1}{N} \sum_{ i, \sigma} \langle
 \, n_{i \alpha, \sigma}  \rangle \,,
    \label{b6}
\end{equation}
where $N$ is the number of unit cells and $\langle ...\rangle$ denotes the statistical
average with the Hamiltonian (\ref{b5}).

\section{Dyson equation}
\label{sec:3}

To consider superconductivity  within the model (\ref{b5}), we introduce the
anticommutator two-time  matrix Green function (GF)~\cite{Zubarev60,Zubarev74}
\begin{eqnarray}
{\sf G}_{i j \sigma} (t-t') =
  - \text{i} \theta(t-t') \langle \{  {\hat X}_{i \sigma }(t) ,\hat{X}^\dag_{j\sigma}(t')\}\rangle
 \equiv \langle \langle {\hat X}_{i \sigma }(t) ,{\hat X}^\dag_{j\sigma}(t')\rangle \rangle\,,
     \label{e1}
\end{eqnarray}
where   $X(t)={\rm e}^{\text{i}Ht} X {\rm e}^{-\text{i}Ht}$ and $\,\theta(x)$ is the Heaviside
function. Here, we use Nambu notation and introduce the Hubbard operators in the
two-sublattice representation  $\hat{X}_{i\sigma }$ and  $\hat{X}^\dag_{i\sigma }$  where
\begin{equation}
 \hat{X}_{i \sigma }= \left( \begin{array}{c} X_{i \text A}^{0 \sigma } \vspace{1mm}\\
 X_{i\text B}^{ 0\sigma} \vspace{1mm}\\  X_{i\text A}^{\bar{\sigma} 0} \vspace{1mm}\\
 X_{i\text B}^{\bar{\sigma} 0} \vspace{1mm}\\
  \end{array} \right),\qquad \hat{X}^\dag_{j\sigma }= \left(  X_{j\text A}^{\sigma 0 }\,
   X_{j\text B}^{\sigma 0} \, X_{j\text A}^{0 \bar{\sigma} }  X_{j\text B}^{0 \bar{\sigma} }  \right) .
  \label{e2}
\end{equation}
The Fourier representation in $({\bf k}, \omega) $-space is defined by
\begin{eqnarray}
{\sf G}_{ij \sigma} (t-t') = \int_{-\infty}^{\infty}\frac{\rd\omega }{2\piup} {\rm e}^{- \textrm
{i}\omega(t-t')} \frac{1}{N}\,
 \sum_{\bf k}{\rm e}^{\textrm
 	{i}{{\bf k} ({\bf r}_{i}-{\bf r}_{j})}} {\sf G}_{\sigma}({\bf k},\omega).
     \label{e1a}
\end{eqnarray}
The $4\times 4$  matrix GF (\ref{e1}) can be written as
\begin{equation}
 {\sf G}_{ij \sigma}(\omega) =\langle\langle {\hat X}_{i\sigma} \mid   \hat
X_{j\sigma}^{\dagger}\rangle\rangle_{\omega} =
  {\hat G_{ij\sigma}(\omega)  \quad \quad
 \hat F_{ij\sigma}(\omega) \choose
 \hat F_{ij\sigma}^{\dagger}(\omega) \quad
   -\hat{G}_{ji\bar\sigma}(-\omega)} ,
 \label{e2a}
\end{equation}
where the normal $\hat G_{ij\sigma}$ and anomalous $\hat F_{ij\sigma}$ components of the
GF are  $2\times 2$ matrices, which are coupled by the symmetry relations for the
anticommutator retarded GF~\cite{Zubarev60,Zubarev74}.

To calculate the GF (\ref{e1}),  we use the equation of motion method. Differentiating
the GF with respect to  time $t$, its Fourier representation leads to the equation
\begin{eqnarray}
 && \omega {\sf G}_{ij\sigma}(\omega) = \delta_{ij} {\sf Q} +
   \langle \langle
    [\hat X\sb{i\sigma},H] \, , \,  \hat X\sb{j\sigma}\sp{\dagger}
   \rangle \rangle_{\omega} \,,
\label{e3}
\end{eqnarray}
where ${\sf Q} = \langle \{ \hat X_{i\sigma},\hat X_{i\sigma}^{\dagger}\}\rangle  =
\tilde{\tau}_{0} Q\, $. Here, $\tilde{\tau}_{0}$ is the $4 \times 4$ unit matrix and in a
paramagnetic state, the coefficient $\, Q  = \langle X\sb{i \alpha}\sp{00} + X\sb{i
\alpha}\sp{\sigma \sigma} \rangle = 1-n_{\alpha}/2\, $ depends  on the occupation number
of electrons  (\ref{b6}) only.
\par
For a system of strongly correlated electrons as in the $t$-$J$ model, there is no
well-defined quasi-particle (QP) excitations specified by zeroth-order kinetic energy.
Therefore, it is convenient to choose the mean-field contribution  to the energy of QPs
in the equations of motion (\ref{e3}) as the zeroth-order QP energy. To identify this
contribution, we use the  projection operator method developed for the GF ~\cite{Plakida11}. To this end, we
write the operator $\hat Z\sb{i\sigma} = [\hat X\sb{i\sigma},H]$ in (\ref{e3}) as a sum
of the linear part, proportional to the original operator $\hat X_{i\sigma}$, and the
irreducible part $\hat Z\sb{i\sigma}\sp{(\text {ir})}$ orthogonal to $\hat X_{i\sigma}$:
\begin{equation}
  \hat Z\sb{i\sigma} = [\hat X\sb{i\sigma}, H] =
    \sum\sb{l}{\sf E}\sb{il\sigma} \hat X\sb{l\sigma} +
    \hat Z\sb{i\sigma}\sp{(\text {ir})}.
\label{e4}
\end{equation}
The orthogonality condition
 $\, \langle \{ \hat Z\sb{i\sigma}\sp{(\text {ir})},
    \hat X\sb{j\sigma}\sp{\dagger} \} \rangle = \langle \hat Z\sb{i\sigma}\sp{(\text {ir})}
    \hat X\sb{j\sigma}\sp{\dagger} +
    \hat X\sb{j\sigma}\sp{\dagger} \hat Z\sb{i\sigma}\sp{(\text {ir})} \rangle = 0 \,$
determines  the linear part, the  frequency matrix:
\begin{eqnarray}
 {\sf E}\sb{ij\sigma}=  \langle \{ [\hat X\sb{i\sigma}, H],
    \hat X\sb{j\sigma}\sp{\dagger} \} \rangle {\sf Q}^{-1} =
 { \hat{E}_{ij} \quad \quad
\hat{\Delta}_{ij\sigma} \choose \hat{\Delta}_{ji\sigma}^{*} \quad -\hat{E}_{ji} }\,
 . \label{e5}
\end{eqnarray}
Frequency matrix (\ref{e5}) determines QP spectrum $\hat{E}_{ij}$  and the gap function
$\hat{\Delta}_{ij\sigma}$ in the  GMFA. The corresponding zeroth-order GF in the Fourier
representation reads
\begin{equation}
  {\sf G}\sp{0}\sb{\sigma }({\bf k},\omega) =
    \bigl[ \omega \tilde \tau\sb{0} - {\sf E}\sb{\sigma}({\bf k})
      \bigr] \sp{-1} {\sf Q} \, , \quad {\sf E}({\bf k})=
 { \hat{E}({\bf k}) \quad \quad
\hat{\Delta}_{\sigma}({\bf k}) \choose \hat{\Delta}_{\sigma}^{*}({\bf k}) \quad
-\hat{E}({\bf k}) }\,  . \label{e6}
\end{equation}
\par
Differentiating the multiparticle GF $\,\langle \langle
    \hat Z\sb{i\sigma} (t) \,  , \,  \hat X\sb{j\sigma}\sp{\dagger}(t')
   \rangle \rangle \,$ in (\ref{e3})  with respect
to the second time $t'$ and using the same projection procedure as in (\ref{e4}) leads to
the Dyson equation for the GF (\ref{e1}). In the $({\bf k}, \omega )$-representation, the
Dyson equation reads
\begin{equation}
  \left[ {\sf G}\sb{\sigma}({\bf k}, \omega) \right]\sp{-1} =
  \left[ {\sf G}\sb{\sigma}\sp{0}({\bf k}, \omega) \right]\sp{-1} -
\widetilde{\Sigma}\sb{\sigma}({\bf k}, \omega). \label{e7}
\end{equation}
The self-energy operator $ \widetilde{\Sigma}_{\sigma}({\bf k},\omega)$ is defined by the
{\it proper} part of the scattering matrix which has no parts connected by the zeroth
order GF (\ref{e6}):
\begin{equation}
 \widetilde{\Sigma}\sb{\sigma}({\bf k}, \omega) = {\sf Q}\sp{-1}
    \langle\langle {\hat Z}\sb{{\bf k}\sigma}\sp{(\text {ir})} \mid
     {\hat Z}\sb{{\bf k}\sigma}\sp{(\text {ir})\dagger} \rangle\rangle
      \sp{(\text{prop})}\sb{\omega}\;{\sf  Q}\sp{-1} .
\label{e8}
\end{equation}
The self-energy operator (\ref{e8}) can be  written in the same form as GF (\ref{e2a}):
\begin{equation}
\widetilde{\Sigma}_{ij\sigma}(\omega) = {\sf Q}^{-1} \, {\hat M_{ij}(\omega) \quad  \quad
\hat\Phi_{ij\sigma}(\omega) \choose \hat\Phi_{ij\sigma}^{\dagger} (\omega)\quad
-\hat{M}_{ji}(-\omega)} {\sf Q}^{-1}  \, .
 \label{e15}
\end{equation}
The normal $\hat M$ and anomalous  (pair) $\hat\Phi$ components of the self-energy
operator (\ref{e15}) are given by the $2\times 2$ matrices [see (\ref{e16}),
(\ref{e17})].

The system of equations for the $(4 \times 4)$ matrix GF (\ref{e2a}) and the self-energy
(\ref{e15}) can be reduced to   a system of equations for the normal ${\hat
G}_\sigma({\bf k},\omega)$ and anomalous ${\hat F}_\sigma({\bf k},\omega)$ $(2 \times 2)$
matrix components.  By using representations for the frequency matrix (\ref{e5}) and the
self-energy (\ref{e15}), we derive    the following system of matrix equations for them:
\begin{eqnarray}
{\hat G}_\sigma({\bf k},\omega)& = &  \Bigl[
  \hat {G}_{N}({\bf k},\omega)^{-1} +
  \hat{\varphi}_\sigma({\bf k},\omega)\,
  \hat{G}_{N}({\bf k},- \omega)\,\hat{\varphi}^{*}_\sigma({\bf
k},\omega)  \Bigr]^{-1}  {Q}\,,
\label{r1} \\
{\hat F}^{\dag}_\sigma({\bf k},\omega)& = & -\hat{G}_{N}({\bf k},-
\omega)\,\hat{\varphi}^{\dag}_\sigma({\bf k},\omega) \,
 \hat{G}_\sigma({\bf k},\omega).
 \label{r2}
\end{eqnarray}
In (\ref{r1}), we introduced  the normal state matrix GF and the matrix superconduction
gap function:
\begin{eqnarray}
{\hat G}_{N}({\bf k},\omega)& = & \Bigl[ \omega \hat\tau_0 - \hat{\varepsilon}({\bf k}) -
  \hat{M}({\bf k},\omega)/ \hat{Q} \Bigr]^{-1},
\label{r3} \\
{\hat \varphi}_\sigma({\bf k},\omega)& = & \hat{\Delta}_{\sigma}({\bf k}) +
 \hat\Phi_{\sigma}({\bf k},\omega) /\hat{Q} .
 \label{r4}
\end{eqnarray}
To obtain a closed system of equations, we should  evaluate the multiparticle GF in the
self-energy operator~(\ref{e15})  which describes the processes of inelastic scattering
of electrons on charge and spin fluctuations due to the kinematic interaction.

\subsection{Mean-field approximation}

The superconducting pairing in the Hubbard model already occurs in the MFA and is caused
by the  kinetic exchange interaction as proposed by Anderson~\cite{Anderson87,Anderson97}. It is,
therefore, reasonable to consider the MFA described by zeroth-order GF (\ref{e6})
separately. Using commutation relations for Hubbard operators~(\ref{b4}), we evaluate the
frequency matrix~(\ref{e5}). The matrix $\hat{E}({\bf k})$ determines the QP spectrum in
the   normal phase (see \cite{Vladimirov18}):
\begin{eqnarray}
  \hat{E}({\bf k})&=&
  \frac{1}{N}\sum_{i,j} \exp[\text {i}{{\bf k} ({\bf r}_i-{\bf r}_j)}] \hat{E}_{ij}
 = \left(
\begin{array}{cc}
\varepsilon_\text{A} ({\bf k})  &  \varepsilon_\text{AB}({\bf k}) \\
     \varepsilon_\text{BA}({\bf k}) & \varepsilon_\text{B} ({\bf k})
\end{array} \right) ,
 \nonumber\\
 \varepsilon_\text A({\bf k})  &= &  \langle \{ [ X_{{\bf k} \text A}^{0 \sigma }, H],
X_{{\bf k} \text A}^{\sigma 0}\} \rangle \; {Q}^{-1} ,
 \quad
 \varepsilon_\text B({\bf k}) =  \langle \{ [ X_{{\bf k} \text B}^{0 \sigma }, H],
X_{{\bf k} \text B}^{\sigma 0}\} \rangle \; {Q}^{-1} ,
 \nonumber\\
 \varepsilon_\text{AB}({\bf k})  &= &  \langle \{ [ X_{{\bf k} \text A}^{0 \sigma }, H],
X_{{\bf k} \text B}^{\sigma 0}\} \rangle \; {Q}^{-1} ,
 \quad
  \varepsilon_\text{BA}({\bf k}) = (\varepsilon_\text{AB}({\bf k}))^\dagger .
\label{e10}
\end{eqnarray}
The solution of the matrix equation for the zero-order GF (\ref{e6}) in the normal state
reads:
\begin{eqnarray}
  \hat G^0_N({\bf k}, \omega) = \left( \begin{array}{cc}
  G^0_\text{AA}({\bf k}, \omega)\quad  G^0_\text{AB}({\bf k}, \omega) \\
    G^0_\text{BA}({\bf k}, \omega)\quad   G^0_\text{BB}({\bf k}, \omega)\\
        \end{array}\right)
      =         \frac{Q}{D({\bf k}, \omega)} \left(
\begin{array}{cc}
\omega - \varepsilon_\text B ({\bf k})   \quad  \varepsilon_\text{AB}({\bf k}) \\
    \varepsilon_\text{AB}({\bf k})^*\quad  \omega- \varepsilon_\text A ({\bf k}) \\
        \end{array}\right) ,
\label{e11}
\end{eqnarray}
where
\begin{eqnarray}
D({\bf k}, \omega) =
  [\omega -\varepsilon_\text A ({\bf k}) ] [\omega -\varepsilon_\text B ({\bf k})  ] -|\varepsilon_\text{AB}({\bf k})|^2
    = [\omega -\varepsilon_{+}({\bf k}) ] [\omega -\varepsilon_{-}({\bf k})  ] .
\end{eqnarray}
The electronic spectrum  has two branches:
\begin{eqnarray}
  \varepsilon_{\pm}({\bf k}) =  \frac{1}{2}\left[\varepsilon_\text A ({\bf k}) +\varepsilon_\text B ({\bf k})\right]
 \pm  \frac{1}{2}\Big\{ \left[\varepsilon_\text A ({\bf k})
- \varepsilon_\text B ({\bf k})\right]^2
   + 4 |\varepsilon_\text{AB}({\bf k})|^2\Big\}^{1/2} = \varepsilon ({\bf k}) \pm |\varepsilon_\text{AB}({\bf k})|\,,
      \label{e12}
\end{eqnarray}
where we take into account the fact that in the paramagnetic state, the sublattices  are equivalent
and, therefore, $\varepsilon_\text A ({\bf k}) = \varepsilon_\text B ({\bf k})= \varepsilon ({\bf k})
$. We calculate the spectrum and consider the Fermi surface in the normal state in the
next section.

Now, we evaluate the anomalous component $\hat{\Delta}_{ij\sigma}$ of the matrix
(\ref{e5}), which determines the superconducting gap. Similar to the normal-state
frequency matrix (\ref{e10}),   we obtain the representation
\begin{eqnarray}
\hat{\Delta}_{\sigma}({\bf k}) &=&
  \frac{1}{N}\sum_{i,j} \exp[\text {i}{{\bf k} ({\bf r}_i-{\bf r}_j)}]\, \left(
\begin{array}{cc}
 \Delta_{i j, \text A \sigma}   & \Delta_{i j, \text{AB} \sigma}  \\
     \Delta_{i j, \text{BA} \sigma}  &  \Delta_{i j, \text B \sigma}
\end{array} \right)
 = \left(
\begin{array}{cc}
\Delta_{\text A\sigma}({\bf k})  &  \Delta_{\text{AB}\sigma}({\bf k}) \\
    \Delta_{\text{BA}\sigma}({\bf k}) &\Delta_{\text B\sigma}({\bf k})
\end{array} \right),
 \nonumber\\
\Delta_{\text A\sigma}({\bf k})  &= &  \langle \{ [ X_{{\bf k} \text A}^{0 \sigma }, H], X_{{\bf k}
\text A}^{0 \bar\sigma }\} \rangle \; {Q}^{-1} ,
 \quad
 \Delta_{\text B\sigma}({\bf k}) =  \langle \{ [ X_{{\bf k} \text B}^{0 \sigma }, H],
X_{{\bf k} \text B}^{0 \bar\sigma } \} \rangle \; {Q}^{-1} ,
 \nonumber\\
 \Delta_{\text{AB}\sigma}({\bf k})  &= &  \langle \{ [ X_{{\bf k} \text A}^{0 \sigma }, H],
X_{{\bf k} \text B}^{0 \bar\sigma } \} \rangle \; {Q}^{-1} ,
 \quad
 \Delta_{\text{BA}\sigma}({\bf k}) =\Delta_{\text{AB}\sigma}^\dagger({\bf k}).
\label{e13}
\end{eqnarray}
Diagonalization of the gap matrix  (\ref{e13}) shows a two-gap structure:
\begin{eqnarray}
 \Delta^{\pm}_{\sigma}({\bf k}) & = &  \frac{1}{2}\big[\Delta_{\text A\sigma} ({\bf k})
  +\Delta_{\text B\sigma} ({\bf k})\big]
 \pm  \frac{1}{2}\Big\{ [\Delta_{\text A\sigma}({\bf k})
- \Delta_{\text B\sigma} ({\bf k})]^2
    + 4 |\varepsilon_\text{AB}({\bf k})|^2\Big\}^{1/2}
    \nonumber\\
    & = & \Delta_{\sigma}({\bf k}) \pm |\Delta_{\text{AB}\sigma}({\bf k})|\,,
      \label{e14}
\end{eqnarray}
where we take into account the fact that in the paramagnetic state, the sublattices  are equivalent
and, therefore, $\Delta_{\text A\sigma}({\bf k}) = \Delta_{\text B\sigma}({\bf k})=
\Delta_{\sigma}({\bf k}) $. We consider the gap components (\ref{e13}) in the next
section.

\subsection{Self-energy operator}

The normal $\hat M$ and anomalous  (pair) $\hat\Phi$ components of the self-energy
operator (\ref{e15}) are given by the $2\times 2$ matrices:
\begin{eqnarray}
 \hat{M}_{ij}(\omega)=
\bigg\langle \bigg\langle
    { [\widetilde{X_{i\text A}^{0\sigma}}, H] \choose
     [\widetilde{X_{i\text B}^{0 \sigma }}, H]}  \mid
 ( [H, \widetilde{X_{j \text A}^{\sigma 0}}]\,
  [H, \widetilde{X_{j \text B}^{\sigma 0}}])
\bigg \rangle \bigg \rangle^{\text{(pp)}}_{\omega}, \quad
\label{e16} \\
\hat{\Phi}_{ij\sigma}(\omega)= \bigg\langle\bigg\langle
    { [\widetilde{X_{i \text A}^{0\sigma}}, H]  \choose
     [\widetilde{X_{i\text B}^{0 \sigma }}, H]} \mid
 ( [H, \widetilde{X_{j\text A}^{0 \bar\sigma }}]
  [H, \widetilde{X_{j\text B}^{0 \bar\sigma}}])
\bigg \rangle \bigg \rangle^{\text{(pp)}}_{\omega}, \quad
 \label{e17}
 \end{eqnarray}
where  $\, [ \widetilde{ X_{i\alpha}^{n, m}} , H] \,$ are  the  irreducible parts of the
commutators  determined by equation~(\ref{e4}). Taking into account  equations  for the $[
X_{i \alpha}^{0 \sigma}, H]$  operators, we obtain the following representation for the
matrix elements of  the normal self-energy
\begin{eqnarray}
 {M}_{ij \text A}(\omega) &= &  t^2 \sum_{ l, \sigma'}\sum_{ k, \sigma''}\,
\langle \langle \widetilde{B}_{i \text A}^{\sigma \sigma'} X_{l \text B}^{0 \sigma'}
    \mid
 X_{k \text B}^{ \sigma'' 0}  \widetilde{B}_{j \text A}^{\sigma'' \sigma}
 \rangle  \rangle^{\text{(pp)}}_{\omega}
 \nonumber \\
&+& \frac{J^2}{4} \sum_{ l, \sigma'}\,\sum_{ k, \sigma''}\, \langle \langle
 \widetilde{B}_{l \text B}^{\sigma \sigma'} X_{i\text A}^{0 \sigma' }
 \mid
 X_{j\text A}^{\sigma'' 0} \widetilde{B}_{k \text B}^{\sigma'' \sigma}
 \rangle  \rangle^{\text{(pp)}}_{\omega},
\label{e16a}
 \end{eqnarray}
\begin{eqnarray}
M_{ij \text{AB}}(\omega)&= &\ t^2 \sum_{ l, \sigma'}\sum_{ k, \sigma''}\, \langle \langle
 \widetilde{B}_{i \text A}^{\sigma \sigma'} X_{l \text B}^{0 \sigma'}
    \mid
 X_{k \text A}^{ \sigma'' 0}  \widetilde{B}_{j \text B}^{\sigma'' \sigma}
 \rangle  \rangle^{\text{(pp)}}_{\omega}
 \nonumber \\
&+& \frac{J^2}{4} \sum_{ l, \sigma'}\,\sum_{ k, \sigma''}\, \langle \langle
\widetilde{B}_{l \text B}^{\sigma \sigma'} X_{i\text A}^{0 \sigma' }
 \mid
X_{jB}^{\sigma'' 0}\widetilde{B}_{k \text A}^{\sigma'' \sigma} \rangle 
\rangle^{\text{(pp)}}_{\omega},
 \label{e16b}
 \end{eqnarray}
where
\begin{eqnarray}
\widetilde{B}_{i\alpha}^{\sigma \sigma'} = ( S_{i\alpha}^z - n_{i\alpha}/2 )\,
\delta_{\sigma ,\sigma'}+ X_{i\alpha}^{{\sigma}' \sigma }\,\delta_{\sigma' ,
\bar{\sigma}}\,, \quad (\widetilde{B}_{i\alpha}^{\sigma \sigma'})^\dag  =
\widetilde{B}_{i\alpha}^{\sigma' \sigma}.
 \label{e16c}
 \end{eqnarray}
In (\ref{e16a}), (\ref{e16b}), we take into account only the diagonal in respect to the
interaction terms. Using equations for the operators $[ X_{i \alpha}^{\bar\sigma 0}, H]$, a
similar representation is derived for the anomalous matrix elements (\ref{e17}):
\begin{eqnarray}
 {\Phi}_{ij \text A}(\omega)=
\langle \langle
   [\widetilde{X_{i \text A}^{0\sigma}}, H] \mid
  [H, \widetilde{X_{j\text  A}^{0 \bar{\sigma} }}]
 \rangle  \rangle^{\text{(pp)}}_{\omega}, \quad
\label{e17a} \\
\Phi_{ij \text{AB}}(\omega)= \langle \langle
   [\widetilde{X_{i\text A}^{0\sigma}}, H] \mid
 [H, \widetilde{X_{j \text B}^{ 0  \bar{\sigma} }}]
 \rangle  \rangle^{\text{(pp)}}_{\omega}  .
\label{e17b}
 \end{eqnarray}
To calculate the self-energy matrix elements in~(\ref{e16a}), (\ref{e16b}), and
(\ref{e17a}), (\ref{e17b}),  we use the SCBA for the multiparticle GFs. For this, it is
convenient to write the self-energy in terms of time-dependent correlation functions. In
particular, for the normal component (\ref{e16a}), we have the spectral representation:
\begin{eqnarray}
&& {M}_{ij \text A}(\omega) =  \sum_{ l, \sigma'}\sum_{ k, \sigma''}\,
 \frac{1}{2\piup}\int_{-\infty}^{\infty}
 \rd z \frac{{\rm e}^{\,\beta z}+1}
  {(\omega - z)}\int_{-\infty}^{\infty}\! \rd t {\rm e}^{\text {i} zt}
 \nonumber   \\
&& \times\; \Big[ t^2 \langle  X_{k\text B}^{ \sigma'' 0} \widetilde{B}_{j \text A}^{\sigma''
\sigma} 
 \widetilde{B}_{i \text A}^{\sigma \sigma'}(t) X_{l \text B}^{0 \sigma'}(t) \rangle
 +  \frac{J^2}{4}  \langle X_{j\text A}^{\sigma'' 0} \widetilde{B}_{k \text B}^{\sigma'' \sigma}
 \widetilde{B}_{l \text B}^{\sigma \sigma'}(t) X_{i\text A}^{0 \sigma' } (t)\rangle \Big].
\label{a19a}
\end{eqnarray}
In the SCBA, a propagation of excitations described by the Fermi-like operators $\,\hat
X_{i}^{ \sigma 0} \,$ and  the Bose-like operators $\widetilde{B}_{j }^{\sigma' \sigma} $
for $i \neq j$ is assumed to be independent.  Therefore, the corresponding time-dependent
multiparticle correlation functions can be written   as a product of fermionic and
bosonic correlation functions,
\begin{eqnarray}
&&  \langle  X_{k \text B}^{ \sigma'' 0} \widetilde{B}_{j \text A}^{\sigma'' \sigma} \;
 \widetilde{B}_{i \text A}^{\sigma \sigma'}(t) X_{l \text B}^{0 \sigma'}(t) \rangle =
\delta_{\sigma\sigma'}\,\langle  X_{k \text B}^{\sigma' 0}  X_{l \text B}^{0 \sigma'}(t) \rangle\;
\langle  \widetilde{B}_{j \text A}^{\sigma' \sigma} \widetilde{B}_{i \text A}^{\sigma \sigma'}(t)
\rangle .
 \label{a20}
\end{eqnarray}
The time-dependent correlation functions in (\ref{a20})  are calculated self-consistently
using the corresponding GFs, as e.g.,
\begin{eqnarray}
\langle  X_{k\text B}^{\sigma' 0}  X_{j\text B}^{0 \sigma'}(t) \rangle & =& \int_{-\infty}^{\infty}
\rd\omega  n(\omega) {\rm e}^{-\text{i}\omega t}
 [-(1/\piup)]\,{\rm Im} \langle \langle  X_{j \text B}^{0 \sigma'}|
 X_{k \text B}^{\sigma' 0}  \rangle  \rangle_{\omega}\,,
 \label{a21a}\\
 \langle  \widetilde{B}_{j \text A}^{\sigma' \sigma} \widetilde{B}_{i\text A}^{\sigma \sigma'}(t)
\rangle
 & = & \int_{-\infty}^{\infty} \rd\omega  N(\omega)
 {\rm e}^{-\text{i}\omega t}
 [-(1/\piup)]\,{\rm Im} \langle \langle
\widetilde{B}_{i\text A}^{\sigma \sigma'}|  \widetilde{B}_{j \text A}^{\sigma' \sigma}
 \rangle  \rangle_{\omega}\,,
 \label{a21b}
\end{eqnarray}
where $n(\omega)$ and  $ N(\omega)$ are the Fermi- and Bose-function, respectively.
Integration over time $t$ in (\ref{a19a})  yields
\begin{eqnarray}
&&{M}_{ij \text A}(\omega) = \sum_{ l, k}\sum_{ \sigma' }\,\int \! \int_{-\infty}^{\infty}
\frac{\rd\omega_1 \rd\omega_2}{\piup^2}
 \; \frac{1 - n(\omega_1)+ N(\omega_2)}
  {\omega -\omega_1 - \omega_2}
\nonumber\\
&& \times\; \Big( t^2 {\rm Im} \langle \langle  X_{l \text B}^{0 \sigma'}|  X_{k \text B}^{\sigma'
0}  \rangle  \rangle_{\omega_1}\, {\rm Im} \langle \langle \widetilde{B}_{i
\text A}^{\sigma \sigma'}|  \widetilde{B}_{j \text A}^{\sigma' \sigma}
 \rangle  \rangle_{\omega_2}
 \nonumber\\
&& +  \frac{J^2}{4}  {\rm Im} \langle \langle  X_{i\text A}^{0 \sigma'}|  X_{j \text B}^{\sigma' 0}
\rangle  \rangle_{\omega_1}\, {\rm Im} \langle \langle \widetilde{B}_{l\text B}^{\sigma
\sigma'}|  \widetilde{B}_{k\text A}^{\sigma' \sigma}
 \rangle  \rangle_{\omega_2}\Big).
 \label{a22}
\end{eqnarray}
Taking into account the definition of the bosonic operator (\ref{e16c}), the bosonic GFs
in these equations can be written as
\begin{eqnarray}
&&\langle \langle \widetilde{B}_{i \text A}^{\sigma \sigma'}|  \widetilde{B}_{j \text A}^{\sigma'
\sigma}
 \rangle  \rangle_{\omega}
= \frac{1}{4}\langle \langle n_{i \text A} | n_{j \text A}\rangle  \rangle_{\omega} \,
\delta_{\sigma'\sigma}
   + \langle \langle  S^z_{i\text A}| S^z_{j\text A}
 \rangle \rangle_{\omega} \, \delta_{\sigma'\sigma}
 + \langle \langle X^{\bar{\sigma}\sigma}_{i\text A}|
 X^{\sigma \bar{\sigma}}_{j\text A}\rangle  \rangle_{\omega}
 \, \delta_{\sigma' \bar{\sigma}} \, .
   \label{a23}
\end{eqnarray}
After summation over $\sigma'$ in equation~(\ref{a22}) for the bosonic GF (\ref{a23}) and
taking into account that the normal GF in the paramagnetic state is independent of spin,
the spin-fluctuation contribution  can be written in the form:
\begin{eqnarray}
\langle \langle  S^z_{i\text A}| S^z_{j\text A}
 \rangle \rangle_{\omega} \,
 + \langle \langle X^{\bar{\sigma}\sigma}_{i\text A}|
 X^{\sigma \bar{\sigma}}_{j\text A}\rangle  \rangle_{\omega}
  =
 \langle \langle {\bf S}_{i\text A}|{\bf S}_{j\text A}
 \rangle \rangle_{\omega} \equiv - \chi_{ij, \text A} (\omega)  .
   \label{a23a}
\end{eqnarray}
Introducing the $\bf k$-representation for the GFs  in the self-energy (\ref{a22}),  we
obtain the expression:
\begin{eqnarray}
{M}_{ \text A}({\bf k},\omega) &= &
   \frac{1}{N} \sum_{\bf q}
   \int\limits\sb{-\infty}\sp{+\infty} \!\!{\rm d}z\,
  \frac{1} {\piup} \int\limits\sb{-\infty}\sp{+\infty}
\rd \Omega \frac{1 +N(\Omega) - n(z)}{\omega - z - \Omega}\,
   \nonumber \\
& \times &  \Big\{ t^2\, \left[-(1/\piup)\right]\,{\rm Im} \langle \langle  X_{\text A{\bf q} }^{0 \sigma}|
X_{\text A{\bf q} }^{\sigma 0}  \rangle  \rangle_{z}\, {\rm Im} \chi_{\text A} ({\bf k -  q},
\Omega)
 \nonumber\\
&& +  \frac{J^2}{4}\, [-(1/\piup)]\, {\rm Im} \langle  \langle  X_{\text A {\bf q} }^{0
\sigma}| X_{\text B{\bf q} }^{\sigma 0}  \rangle  \rangle_{z} {\rm Im} \chi_\text{BA} ({\bf k -
q}, \Omega) \Big\}.
 \label{a24}
 \end{eqnarray}
 The SCBA for ${M}_{ij \text{AB}}(\omega)$  (\ref{e16b}) and for the anomalous self-energies
(\ref{e17a}), (\ref{e17b}) gives similar results.

\section{Results}
\label{sec:4}

To find out   the normal state GF (\ref{r3}) beyond the MFA, the self-energy matrix
elements (\ref{e16a}), (\ref{e16b}) should be calculated. Similar calculations of the 
anomalous self-energy (\ref{e17a}), (\ref{e17b})
 give an equation for the frequency dependent gap (\ref{r4}). However,
implementation of this program demands  complicated self-consistent numerical
computations for the two-band $t$-$J$ model which are much more complicated than
similar calculations performed for the one-band $t$-$J$ model on the square lattice
in~\cite{Plakida99}. Therefore, in the present paper, we restrict our  consideration only
to the results in MFA and study the complete system of equations later.

\subsection{Electronic spectrum}

\begin{figure}[!b]
\centerline{\includegraphics[width=0.4\textwidth]{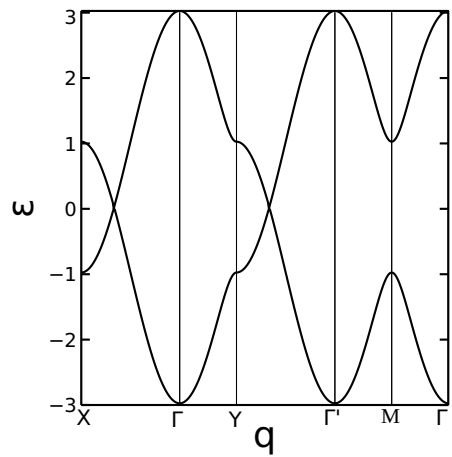}} 
\caption{ Electronic
spectrum $\varepsilon_{\pm}({\bf k})$ in units of $\widetilde{t}$
 at $\tilde{\mu} = 0$.}
\label{fig1}
\end{figure}

\begin{figure}[!t]
\centerline{\includegraphics[width=0.5\textwidth]{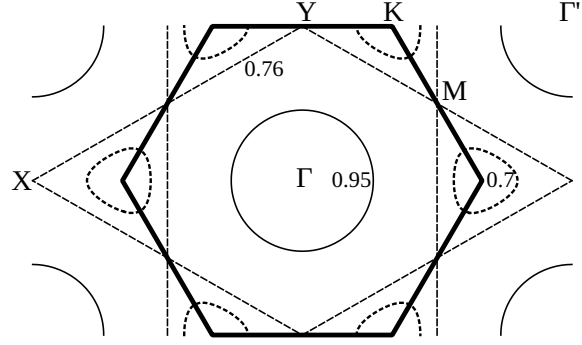}}
 \caption[]{Brillouin zone (bold) and hole  Fermi surface for  $n = 0.95$
(thin solid), $0.76$ (dashed), and $0.7$ (dotted).}
 \label{fig2}
\end{figure}

Calculation of the matrix elements (\ref{e10}) gives the following results:
\begin{eqnarray}
&&  \varepsilon({\bf k})  =
  - \mu +   \frac{3t}{Q}\, D_1 - \frac {3J}{4} n_\alpha +
\frac {3J}{2Q}\, C_1
 \equiv -\widetilde{ \mu }\,, \; \label{12a}
\end{eqnarray}
where we introduce the nn correlation functions   for electrons and spins,
\begin{eqnarray}
D_1 = \langle X^{\sigma 0}_{i\text A} X^{0\sigma}_{i+\delta_1, \text B}\rangle\,, \quad C_1 = \langle
S^z_{i\text A} {S}^z_{i+\delta_1, \text B} \rangle . \label{m7}
\end{eqnarray}
For the off-diagonal energy, we have:
\begin{eqnarray}
&&\varepsilon_\text{AB}({\bf k}) = - \widetilde{t}\, \gamma_1 ({\bf k})\,, \quad
  \widetilde{t}  =  t Q \left(1 + \frac{3 C_1}{2Q^2}\right) + J\frac { D_1}{2 Q}\,,
\label{12b}
\end{eqnarray}
where $\gamma_1({\bf k}) = \sum_\alpha \exp (\text{i} {\bf k} \boldsymbol\delta_\alpha)$ and $
|\gamma_1({\bf k})|^2 = {1 + 4 \cos ( \sqrt{3}k_x /2)[\cos (\sqrt{3} k_x /2) + \cos
({3}k_y/{2})]} $. Thus, the electronic spectrum has two branches
\begin{eqnarray}
\varepsilon_{\pm}({\bf k}) = -\mu \pm  \widetilde{t}\,|\gamma_1({\bf k})|\,, \label{12c}
\end{eqnarray}
which is similar to the spectrum of the graphene (see, e.g., reference~\cite{Neto09}) but with
the renormalized chemical potential $\widetilde{ \mu }$ due to strong correlations,
equation~(\ref{12a}),  and the hopping parameter  $\widetilde{t}\,$,   equation~(\ref{12b}).
Therefore, the spectrum shows a cone-type behaviour at Dirac points $K$ and $K'$ at
the corners of the graphene Brillouin zone (BZ) as shown in figure~\ref{fig1}. In 
figure~\ref{fig2}, the BZ and the Fermi surfaces~(FS) are shown for holes  at the electronic
occupation numbers $n = 0.95$,  $0.76,$ and $0.7$. At $n \lesssim 1$, the hole FS is
small and centered at the $\Gamma$ point. With a decreasing $n$, the FS becomes larger, and
at some characteristic value  $n_0 = 0.76 $, the  FS touches the BZ at M-points. At a larger
hole doping,  six  pockets centered at the $K$-points emerge which shrink to points for
the half-filled band at $n =2/3$.

\par
For the diagonal  GF, we have
\begin{eqnarray}
G_{\alpha \alpha, \sigma}({\bf k},\omega)
 = \frac{Q}{2 }\,\left[ \frac{ 1}
 {\varepsilon_{+}({\bf k}) - \omega}  + \frac{1}
{\varepsilon_{-}({\bf k}) - \omega}\right] . \label{e10a}
\end{eqnarray}
The average occupation number of electrons in one sublattice is equal to
\begin{eqnarray}
n_{\alpha \sigma}({\bf k}) &=&\langle X_{{\bf k}\alpha }^{\sigma \sigma} \rangle =
\langle X_{{\bf k}\alpha }^{\sigma 0}  X_{{\bf k}\alpha }^{0 \sigma} \rangle
\nonumber\\
 &=&\int^{\infty}_{-\infty} \frac{\rd\omega}{{\rm e}^{\omega/T} +1}\, \left[-\frac{1}{\piup}\right] \mbox{Im}
G_{\alpha \alpha, \sigma}({\bf k},\omega)\,,
 \label{e11a}
\end{eqnarray}
which yields in MFA
\begin{equation}
  n_{\alpha,\sigma}({\bf k}) = Q \,\frac{1}{2}[N(\varepsilon_{+}({\bf k}))+ N(\varepsilon_{-}({\bf k}))],
 \label{e11b}
\end{equation}
where $  N(\varepsilon_{\pm}({\bf k})) = \{\exp[\varepsilon_{\pm}({\bf k})/T] +1\}^{-1} $,
and the average number of electrons in one sublattice is given by
\begin{eqnarray}
 n_{\alpha} = \frac{2}{N}\sum_{{\bf k}} n_{\alpha \sigma}({\bf k})
 \label{e11c}
\end{eqnarray}
with $ n_{\alpha}  \leqslant 1$. For the off-diagonal GF in equation~(\ref{e8}), we obtain
\begin{eqnarray}
&& G_{\text{AB}, \sigma}({\bf k},\omega) = Q \frac{\varepsilon_\text{AB}({\bf k})}{2\,
|\varepsilon_\text{AB}({\bf k})|}
 \Big[\frac{1}{\varepsilon_{+}({\bf k}) - \omega} - \frac{1}{\varepsilon_{-}({\bf
k}) - \omega}\Big].
 \label{e10b}
\end{eqnarray}
For the  nn correlation function, we get
\begin{eqnarray}
&&  \langle X_{{\bf k} \text B }^{\sigma 0}  X_{{\bf k} \text A}^{0 \sigma} \rangle
 =  Q\,\frac{\gamma_1({\bf k})}
   {2|\gamma_1({\bf k})|}
 [  N(\varepsilon_{-}({\bf k}))  - N(\varepsilon_{+}({\bf k}))].
 \label{e12a}
\end{eqnarray}
These results can be used in calculating the self-energy in the normal state
(\ref{a24}).

\subsection{Gap equation}

The matrix elements of the gap function (\ref{e13})  are given by
\begin{eqnarray}
\Delta_{i j, \text A \sigma}  &= &  \langle \{ [ X_{i \text A}^{0 \sigma }, H], X_{j \text A}^{0
\bar{\sigma}}\} \rangle \; {Q}^{-1} , \quad \Delta_{i j, \text{AB} \sigma} =  \langle \{ [ X_{i
\text A}^{0 \sigma }, H], X_{j\text B}^{ 0 \bar{\sigma}} \} \rangle \; {Q}^{-1} .
 \label{a10}
\end{eqnarray}
Using the commutation relations for the HOs, for these functions we obtain  the following
representation:
\begin{eqnarray}
\Delta_{i j, \text A \sigma}  &= &   - t \sum_{ k, \sigma'}\, \langle \{ B_{i \text A}^{\sigma
\sigma'} X_{k \text B}^{0 \sigma'}, X_{j \text A}^{0 \bar{\sigma}}\} \rangle {Q}^{-1}
+\frac{J}{2}  \sum_{ k, \sigma'}\,\langle
 \{ (B_{k\text B}^{\sigma \sigma'} - \delta_{\sigma ,\sigma'} ) X_{i\text A}^{0 \sigma' } ,
   X_{j \text A}^{0 \bar{\sigma}} \}\rangle {Q}^{-1}
 \nonumber\\
&= & - t\delta_{i, j} \sum_{ k}\,\langle X_{i \text A}^{ 0 \sigma} \, X_{k \text B}^{0\bar{\sigma}} -
   X_{i \text A}^{0 \bar{\sigma} }X_{k \text B}^{0 \sigma}\rangle {Q}^{-1} = \delta_{i, j} 2t \sum_{ \delta_\alpha}\,\langle
   X_{i \text A}^{0 \bar{\sigma} }X_{i+ \delta_\alpha \text B}^{0 \sigma}\rangle  {Q}^{-1},
 \label{a10a}
\end{eqnarray}
\begin{eqnarray}
\Delta_{i j, \text{AB} \sigma}  &= & \frac{J}{2}  \sum_{ k, \sigma'}\,
\delta_{k,j}\left(\delta_{\sigma' ,\bar{\sigma}}
   \langle X_{j\text B}^{0 {\sigma}}   X_{i\text A}^{0 \bar{\sigma}}\rangle -
  \delta_{\sigma' ,{\sigma}} \langle X_{j\text B}^{0 \bar{\sigma}}   X_{i\text A}^{0 {\sigma}} \rangle\right) {Q}^{-1}
 \nonumber\\
   & = & - J \, \langle  X_{i\text A}^{0 \bar{\sigma}} X_{j \text B}^{0{\sigma}}  \rangle \; {Q}^{-1},
   \label{a11a}
 \end{eqnarray}
where the symmetry relations were used: $\, \langle X_{j \text B}^{0 {\sigma}}   X_{i\text A}^{0
\bar{\sigma}}\rangle = - \langle  X_{i\text A}^{0 \bar{\sigma}} X_{j \text B}^{0 {\sigma}}  \rangle
\,$  and  $\, \langle X_{j\text B}^{0 \bar{\sigma}}   X_{i\text A}^{0 {\sigma}} \rangle  = - \langle
X_{j \text B}^{0 {\sigma}}   X_{i\text A}^{0 \bar{\sigma}} \rangle\,$.

 Fourier transformation results in the gap function components:
 \begin{eqnarray}
  \Delta_{\text A \sigma}({\bf k})&=&
  2t\,  \sum_{\delta_\alpha}\,\langle
   X_{i \text A}^{0 \bar{\sigma} }X_{i+ \delta_\alpha \text B}^{0 \sigma}\rangle  {Q}^{-1},
 \label{a12a}\\
  \Delta_{\text{AB} \sigma}({\bf k})&=&
  - J \sum_{\delta_\alpha}\, \exp[\text{i}{{\bf k} \boldsymbol\delta_\alpha}]\
\langle  X_{i\text A}^{0 \bar{\sigma}} X_{i+ \delta_\alpha \text B}^{0{\sigma}}  \rangle  {Q}^{-1}
, \label{a12b}
\end{eqnarray}
where $ {\boldsymbol \delta}_\alpha = \delta_1, \delta_2, \delta_3 $. Introducing the
bond-dependent anomalous correlation functions
\begin{eqnarray}
F^{\alpha}_{\text{AB} \sigma} = \langle  X_{i\text A}^{0 \bar{\sigma}}
   X_{i+ \delta_\alpha \text B}^{0{\sigma}} \rangle\,,
\label{a12c}
\end{eqnarray}
the gaps can be written as
\begin{eqnarray}
  \Delta_{\text A \sigma}=
   2t\,  \sum_{\delta_\alpha}\,F^{\alpha}_{\text{AB} \sigma} {Q}^{-1}, \quad
   \Delta_{\text{AB} \sigma}({\bf k}) =
   - J \sum_{\delta_\alpha}\, \exp[\text{i}{{\bf k} \boldsymbol\delta_\alpha}]\, F^{\alpha}_{\text{AB} \sigma}
  {Q}^{-1} .
\label{a13}
\end{eqnarray}
Here, the nn correlation function (\ref{a12c})  is calculated using the spectral
representation:
\begin{eqnarray}
\langle  X_{i\text A}^{0 \bar{\sigma}}    X_{j \text B}^{0{\sigma}} \rangle = \int_{-\infty}^{\infty}
\rd\omega  n(\omega)
 [-(1/\piup)]\,{\rm Im} \langle \langle  X_{j \text B}^{0{\sigma}}|
 X_{i\text A}^{0 \bar{\sigma}}   \rangle  \rangle_{\omega}\,.
 \label{a13b}
\end{eqnarray}
For the bond-independent correlation functions ($s$-wave pairing), $ F^{\alpha}_{\text{AB}
\sigma} = F_{\text{AB} \sigma}$, we have the results similar to the normal state energy:
\begin{eqnarray}
  \Delta_{\text A \sigma}=
   6t F_{\text{AB} \sigma} \; {Q}^{-1},  \quad
  \Delta_{\text{AB} \sigma}({\bf k}) =
   - J \gamma({\bf k}) F_{\text{AB} \sigma}  \; {Q}^{-1} .
\label{a14}
\end{eqnarray}
Using the GMFA for the anomalous GF in (\ref{a13b}), we can derive the Bardeen-Cooper-Schrieffer (BCS) type equations as was proposed in several publications for   electrons with the Dirac spectrum. In particular, in~\cite{Kopnin08}, the $s$-wave BCS model was considered with a phenomenological coupling constant.  In the case of the two-subband honeycomb lattice, we expect to obtain a more complicated pairing symmetry, e.g., $d + \ri d$ singlet pairing or $p$ triplet pairing   as proposed in~\cite{Gu13}. In the case of the $t$-$J$ on the square lattice in~\cite{Plakida99}, we have also found a more complicated   $d$-wave pairing induced by spin-fluctuations. A self-consistent solution of the gap equations  (\ref{a13}) with  microscopic coupling constants  and  anomalous GF (\ref{r2}) and a resulting
equation for determining the superconduction transition temperature will be considered
in a subsequent  publication.

\section{Conclusions}
\label{sec:5}

In the present paper, a microscopic  theory of the electron spectrum and superconductivity
within the two-band $t$-$J$ model for strongly correlated electrons on the honeycomb
lattice  is presented. Using the  projection operator method for 
thermodynamic GFs, we  derived a self-consistent system of equations for the matrix GF
and the self-energy in the SCBA. The latter is similar to the Migdal-Eliasberg
strong-coupling theory for the electron-phonon system. In the present paper, we consider
only the results obtained in the MFA.
However, as it was  pointed out in the  microscopic theory of superconductivity
formulated for the Hubbard model on the square lattice in~\cite{Plakida13,Plakida14},
consideration of the self-energy effects is very important. The normal self-energy
component determines a reduction of the QP weight which can be quite large. The anomalous
(pair) self-energy component strongly enhances the superconduction  pairing induced by
the kinematical interaction of electrons with spin-fluctuations. This interaction being
proportional to the hopping parameter $t$ [see equation~(\ref{a24})]  is  an order
of magnitude larger than the
exchange interaction $J$ considered  in the Anderson's theory of the resonating valence
bonds~\cite{Anderson87,Anderson97}. We are planning to take into account the self-energy effects in
a subsequent publication.

\section*{Acknowledgements}

The  author would like to thank A.A.~Vladimirov and D. Ihle for valuable discussions.

\vspace{-4mm}

\ukrainianpart 

\title{Електронний спектр та надпровідність в $t$-$J$ моделі на стільниковій 
	ґратці} 
\author{Н.М. Плакіда} 
\address{Об'єднаний інститут ядерних досліджень, 141980 Дубна, Росія} 

\makeukrtitle 

\begin{abstract} 
	\tolerance=3000%
	Сформульована мікроскопічна теорія електронного спектра та надпровідності в 
	рамках $t$-$J$ моделі на стільниковій ґратці. З допомогою техніки проекційних 
	операторів Морі отримано рівняння Дайсона для нормальних і аномальних функцій 
	Ґріна для двозонної моделі в термінах операторів Хаббарда. Власноенергетична 
	частина розрахована у самоузгодженому борнівському наближенні для розсіяння 
	електронів на спінових і зарядових флуктуаціях індукованих кінематичною 
	взаємодією для операторів Хаббарда. Обговорюється можливість надпровідного 
	спарювання через антиферомагнітний обмін та спінові флук\-ту\-ації. 
	\keywords сильні електронні кореляції, $t$-$J$ модель, стільникова ґратка, 
	надпровідність 
	
\end{abstract} 

\end{document}